\documentclass[10pt]{iopart}
\input psfig.sty

\begin{document}
\title{Freezing and correlations in fluids with competing interactions} 

\author{D~Pini\S, A~Parola\# and L~Reatto\S}

\address{\S \ Dipartimento di Fisica, Universit\`a di Milano, 
Via Celoria 16, 20133 Milano, Italy}  
\address{\#\ Dipartimento di Scienze Fisiche, 
Universit\`a dell'Insubria, Via Valleggio 11, 22100 Como, Italy}

\ead{davide.pini@mi.infm.it} 

\begin{abstract}

We consider fluids where the attractive interaction at distances slightly
larger than the particle size is dominated at larger distances by a repulsive
contribution. A previous investigation of the effects of the  
competition between attraction and repulsion on  
the liquid-vapour transition and on the correlations 
is extended to the study of the stability of 
liquid-vapour phase separation with respect to freezing. We find that 
this long-range repulsive part of the interaction expands the region where
the fluid-solid transition preempts the liquid-vapour one, so that the critical
point becomes metastable at longer attraction ranges
than those required for purely attractive potentials. 
Moreover,
the large density fluctuations that occur  
near the liquid-vapour critical point are greatly enhanced
by the competition between attractive and repulsive forces, and encompass
a much wider region than in the attractive case. 
The decay of correlations for states where the compressibility is large
is governed by two characteristic lengths, and the usual Ornstein-Zernike
picture breaks down except for the very neighborhood of the critical point,
where one length reduces to the commonly adopted correlation length, while
the other one saturates at a finite value.  

\end{abstract}

\section{Introduction}
\label{sec:introduction}

The attraction between the particles of a fluid is generally the dominant 
contribution to the overall interaction at distances slightly larger than
the particle size $\sigma$. When a longer-ranged repulsion is also present,
the interaction turns from attractive to repulsive as the distance is 
increased beyond $\sigma$, resulting in the so-called {\em mermaid potential} 
(attractive head and repulsive tail). It is well known that, even though 
the integrated intensity of the interaction remains negative, the phase 
diagram of the fluid can be deeply affected by the presence of a repulsion 
at long distance. In fact, when the repulsion strength is large enough,  
the liquid-vapour transition that would take place in the purely attractive
case is inhibited, because the creation of a bulk dense phase becomes 
energetically unfavored. As the temperature is lowered, the fluid particles
instead arrange themselves into a nonhomogeneous 
state~\cite{brazovskii,andelman} with a modulated density, typically 
a cluster- or stripe-shaped configuration. In off-lattice systems like that 
considered here, this phenomenon has been 
observed experimentally in a number of cases~\cite{seul}, including 
colloidal films of metallic particles coated by a surfactant~\cite{gelbart1},
and has been investigated by numerical simulations in two-dimensional model 
fluids~\cite{gelbart2,imperio}.     
The competing interaction scenario has recently gained much attention
because of its relevance in colloidal systems, where the repulsion may be due
to electrostatic interactions or nonideal depletant agents. Cluster formation
induced by competing interactions is frequently found in this context, and 
has been detected for instance in colloid-polymer
mixtures~\cite{bartlett,stradner}, star-linear polymer 
mixtures~\cite{stiakakis}, and protein solutions~\cite{stradner}, 
and can play an important role even 
in the dynamical arrest transition~\cite{sciortino}.  

On the other hand, the competition between the attractive and the repulsive
part of the potential strongly influences the behaviour of the system even 
in the regime where the liquid-vapour transition has not been replaced yet 
by the occurrence of a nonhomogeneous state. In a previous 
study~\cite{chemlet}, we focused on this situation by considering a model 
fluid with a two-body potential given by the superposition of an attractive 
Yukawa tail with inverse-range parameter $z_{1}$ and a longer-ranged, 
repulsive one with inverse-range parameter $z_{2}$, and a strength not large
enough to cause microphase separation. In that work we  
used liquid-state 
theory to study the thermodynamics and the correlations of the fluid
when the amplitude of the repulsion is increased
so as to approach the stability limit of the liquid-vapour transition. 
We found that there are three main effects that occur close to this limit:
i) the size of the region in the temperature-density plane where 
the compressibility is large is remarkably bigger than the near-critical 
region of a fluid with purely attractive interaction; ii) the top of 
the liquid-vapour coexistence curve becomes very flat, so that the transition
is nearly first order; iii) the two-body correlations 
in the region of large compressibility have a faster decay at long distance 
than in the purely attractive case, but on the other hand they are 
strongly enhanced at short and intermediate distance, ranging from near 
contact up to several times and even several tens of the particle size. 
This behaviour indicates that the density fluctuations 
responsible for the large values of the compressibility must be traced back
to the appearance of delocalized clusters of strongly correlated particles.
These can be considered as the precursors of the nonhomogeneous phases that 
eventually set in for strong enough repulsion. The correlation function can 
then be approximately split into an intra-cluster and an inter-cluster
contribution, each governed by a characteristic length. 

In the aforementioned investigation we were almost exclusively concerned 
with the liquid-vapour transition. Little attention was paid to the freezing
transition, whose location was just estimated by means of the Hansen-Verlet
criterion~\cite{verlet} for a specific choice of the inverse-range  
parameters $z_{1}=1$, $z_{2}=0.5$ in units of $\sigma^{-1}$. In that case 
the freezing line was confined
in the high-density region, well beyond the liquid-vapour critical point, 
and, unlike the liquid-vapour transition, it did not appear to be qualitatively 
affected by the competition between attraction and repulsion. 

In view of the rekindled interest in these systems, a more thorough 
investigation is worth being pursued. Therefore, in the present paper we have
extended our previous study by considering 
the freezing and melting transitions for a wide choice of the range 
of the interatomic interaction. Our aim in doing this is twofold:
on the one hand, we would like to determine when the liquid-vapour transition 
is stable with respect to freezing, so that the modification in the shape 
of the liquid-vapour coexistence curve due to the competing interactions can 
actually be observed. On the other hand, we are also interested in finding out
whether the freezing transition itself can be affected by this competition.     To clarify these points, we have employed the self-consistent Ornstein-Zernike
approximation (SCOZA) for the fluid, which we already used in our previous 
work~\cite{chemlet}, and supplemented it by standard thermodynamic 
perturbation theory for the solid phase. 
Compared to the semi-empirical
freezing criteria, which are based on indicators of the transition purely
in the fluid phase, either structural (Hansen-Verlet~\cite{verlet}) 
or entropic (Giaquinta-Giunta~\cite{giaquinta}), this approach has 
the advantage of yielding both the freezing and the melting line, and above 
all it is expected to remain reliable even for interactions with narrow 
attractive parts. 

According to our results, in most cases the stability of the liquid-vapour 
transition with respect to freezing is determined by the inverse range $z_{1}$
of the attractive part of the potential: when $z_{1}$ is small enough,
or conversely large enough, the qualitative shape of the phase diagram is 
basically the same we would expect for a purely attractive interaction with
the same value of $z_{1}$, showing respectively a stable or a metastable 
liquid-vapour transition. However, there exists an intermediate interval 
of $z_{1}$, roughly between $z_{1}=3$ and $z_{1}=6$ in units of $\sigma^{-1}$, 
where the repulsive 
contribution to the interaction strongly influences the location of the 
liquid-vapour transition with respect to the fluid-solid one: while in 
the purely attractive case this range of $z_{1}$ yields stable 
liquid-vapour phase
separation, we find that the latter is preempted by freezing for suitable 
values of the amplitude $A$ and inverse range $z_{2}$ of the repulsion. 
Although this regime does not cover a very wide parameter range, it is 
nevertheless interesting to consider it, since when it occurs, the fluid-solid
transition is affected by the peculiar behaviour of the thermodynamics 
and correlations already discussed in reference~\cite{chemlet} and summarized 
in points i) \ldots iii) above. Specifically, one observes large values of 
the compressibility and very strong correlations along a wide portion of 
the liquidus line. The enhancement of the compressibility is considerably 
stronger than that found in fluids with short-ranged, purely attractive 
interactions 
when the metastable liquid-vapour critical point lies in the neighborhood 
of the liquidus line. 

A second point we want to address in the present paper is a more thorough 
investigation of the behaviour of the correlation function and of the structure
factor $S(q)$.
In fluids with purely attractive tail potentials,
the long-range behaviour of the correlations in the region where 
the compressibility is large is well described in the Ornstein-Zernike (OZ)
picture by an exponential decay with a certain characteristic length, namely
the usual correlation length. On the other hand, when competing interactions
are present, the OZ form is of limited use, save for 
the very neighborhood of the critical region. 
In general, a double-exponential
tail with two characteristic lengths is necessary in order to model the decay
of the correlations, and $1/S(q)$ has to be represented by terms 
up to $q^{4}$.   

The paper is organized as follows: in Sec.~\ref{sec:theory} we give a brief 
sketch of SCOZA and we review the perturbation theory for the solid phase
that we used together with SCOZA to obtain the fluid-solid equilibrium lines.
In Sec.~\ref{sec:freezing} we discuss our results for the stability of 
the liquid-vapour transition with respect to freezing and how the latter 
is affected by the competing interactions. In Sec.~\ref{sec:correlations} 
we focus on the behaviour of the correlations. 
In Sec.~\ref{sec:conclusions} we draw our conclusions.        
 
\section{Theory}
\label{sec:theory}
 
Our model interaction is the hard-core plus two-Yukawa (HCTY) tail potential
that we already considered in reference~\cite{chemlet}:
\begin{equation} 
\fl v(r)=
\left\{
\begin{array}{ll}
\infty \hspace*{5.0cm} & r \le \sigma \, , \\
\epsilon\sigma\left\{\displaystyle{-\frac{1}{r}}\exp[-z_{1}(r/\sigma-1)]+
\displaystyle{\frac{A}{r}}\exp[-z_{2}(r/\sigma-1)]\right\} 
& r > \sigma \, .
\end{array}
\right.
\label{pot}
\end{equation}    
Here $\sigma$ is the hard-sphere diameter, $\epsilon$ is the interaction 
strength, $z_{1}$, $z_{2}$ are the inverse range parameters of the attractive
and repulsive contribution respectively, and $A>0$ is the relative amplitude
of the repulsive contribution. In the following we will always assume 
$z_{1}>z_{2}$, so that $v(r)$ is attractive at distances slightly larger than
$\sigma$, but becomes repulsive at long distance. The SCOZA for the HCTY 
potential has already been described in reference~\cite{scozatwoyuk}, 
so a detailed
derivation of the theory will be omitted here. We just recall that this 
approach can be regarded as a generalization of the mean spherical 
approximation (MSA)~\cite{hansen}. In both MSA and SCOZA, 
the spatial dependence 
of the direct correlation function $c(r)$ for $r>\sigma$ is assumed to be 
the same as that of the tail potential $v(r)$. However, in SCOZA the amplitude
$K$ of $c(r)$ is regarded as an {\em a priori} unknown function 
of the thermodynamic
state, which must be determined by imposing the consistency between 
the compressibility and the internal energy route to thermodynamics. 
This constraint is embodied in the condition:
\begin{equation}
\frac{\partial}{\partial \beta}\left(\frac{1}{\chi_{\rm red}}\right)=
\rho \frac{\partial^2 u}{\partial \rho^2} \, ,
\label{consist}
\end{equation}
where $\rho$ is the number density, $\beta=1/(k_{\rm B}T)$ is the inverse
temperature, $\chi_{\rm red}$ is the reduced compressibility determined
as the structure factor $S(k)$ at $k=0$, and $u$ is the excess 
internal energy per unit volume determined as the spatial integral 
of the potential $v(r)$ weighted by the radial distribution function $g(r)$. 
When the expressions for $\chi_{\rm red}$ 
and $u$ in terms of $c(r)$ and $g(r)$ are substituted 
into Eq.~(\ref{consist}), one obtains a partial differential equation (PDE) 
for the amplitude $K(\rho, \beta)$, which is solved numerically. When the tail
potential is given by a one- or two-Yukawa form, or even by a linear 
superposition of Yukawas, one can take advantage of a wealth of analytical
results~\cite{twoyuk} for the Ornstein-Zernike equation relating 
$c(r)$ to $h(r)\equiv g(r)-1$, and this considerably simplifies 
the solution procedure.     

As stated in the Introduction, the main purpose of the present investigation
of the mermaid potential~(\ref{pot}) is the determination not only 
of the liquid-vapour phase boundary as already done 
in reference~\cite{chemlet},
but also of the fluid-solid one, for which we need the free energies of both
the fluid and solid phases. While the free energies of the liquid 
and the vapour have been obtained by SCOZA, for the free energy of the solid
we have resorted to standard thermodynamic perturbation theory.  
This can be regarded as the application to a solid reference system 
of the perturbative expansion in powers of the inverse temperature $\beta$ 
originally developed for a fluid by Barker and Henderson~\cite{barker}. 
The Helmholtz free energy per particle $F_{s}/N$ of a solid of $N$ particles
interacting via a hard-sphere plus tail potential then reads:
\begin{equation}
\frac{\beta F_{s}}{N}=\frac{\beta F_{s}^{\rm HS}}{N}+2\pi\beta\rho
\int_{0}^{\infty} \!\!dr \, r^{2}g_{s}^{\rm HS}(r) \, v(r) + 
\frac{\beta F_{2}}{N}+{\cal O}(\beta^{3})\, ,
\label{solid}
\end{equation}       
where $F_{s}^{\rm HS}$ and $g_{s}^{\rm HS}$ are the Helmholtz free energy 
and the radial distribution function of the hard-sphere solid averaged over
the solid angle, and $\beta F_{2}/N$ denotes the second-order term in $\beta$. 
Solid-state perturbation theory has been adopted for studying fluid-solid 
equilibrium~\cite{weis} and how it is affected by the range of
the interaction~\cite{gast,hagen,dijkstra}. This approach lacks the flexibility 
of density-functional theory, since it assumes that the perturbed system has 
the same crystal structure as the hard-sphere solid. Nevertheless, when this
is the case, use of the truncated expansion~(\ref{solid}) for the solid phase
has proved to be quite reliable, actually considerably more so than in 
the fluid region~\cite{dijkstra}. In reference~\cite{foffi}, Eq.~(\ref{solid})
was used together with SCOZA to study the phase diagram of a hard-core fluid
with a narrow attractive Yukawa tail, and the results were found to be
very satisfactory. Following the treatment given there, Eq.~(\ref{solid}) 
has been truncated at the second-order term in $\beta$, which has then been 
estimated by an approximation suggested by Barker and Henderson~\cite{barker}, 
namely:   
\begin{equation}
\frac{\beta F_{2}}{N}\simeq
-\pi\beta^{2}\rho\chi_{s}^{\rm HS}
\int_{0}^{\infty} \!\! dr\, r^{2} g_{s}^{\rm HS}(r)\, v^{2}(r) \, ,
\label{mc}
\end{equation}
where $\chi_{s}^{\rm HS}$ is the compressibility of the hard-sphere 
solid divided by the ideal-gas value. The Helmholtz free energy of 
the hard-sphere crystal needed in Eq.~(\ref{solid}) has been determined
by integrating with respect to density the equation of state (EOS) given 
by Hall~\cite{hall}, which expresses the compressibility factor
$Z=\beta P/\rho$ of the hard-sphere solid as an expansion in powers 
of the relative deviation of the density from its value at close packing 
$\rho_{\rm cp}=\sqrt{2}\sigma^{-3}$. As in reference~\cite{foffi}, 
the integration
constant is given by the value of $F_{s}^{\rm HS}$ at a density 
$\rho=0.736\, \rho_{\rm cp}$, for which we have used the result for the excess 
free energy with respect to the ideal gas 
$\beta F_{ex}^{\rm HS}/N=5.91889$, also reported in reference~\cite{foffi},
determined via numerical simulation by Frenkel. The radial distribution 
function of the hard-sphere solid has been represented by the parameterization 
of Kincaid and Weis~\cite{kincaid} with 60 neighbor 
shells~\cite{hirschfelder}. This corresponds to a maximum 
interparticle separation slightly larger than 8 times 
the nearest-neighbor distance $d$, which in turn is at most about 
$d\simeq 1.125 \sigma$ in the density range for the solid phase that we
investigated. In practice, we cut off the integrals appearing 
in Eqs.~(\ref{solid}), (\ref{mc}) at $r_{0}=8\sigma$, and then we added 
a long-range correction by setting $g_{s}^{\rm HS}(r)\equiv 1$ for $r>r_{0}$.   Such a correction is important only for the longest-ranged potentials among 
those considered here. 
Fluid-solid equilibria were determined by equating the pressure and 
the chemical potential of the fluid phase determined by SCOZA with those
of the solid phase given by Eqs.~(\ref{solid}), (\ref{mc}). 

It should be noted
that for this procedure to be a viable one, it is essential that 
the reference hard-sphere system is described at the same level of accuracy 
in both phases. Specifically, use of the accurate Hall EOS for the hard-sphere
solid requires that the Carnahan-Starling (CS) equation~\cite{hansen}, 
or one of comparable
accuracy, is used for the fluid. When the CS EOS is used, the hard-sphere 
freezing and melting densities are found to be $\rho_{f}=0.942\sigma^{-3}$ 
and $\rho_{m}=1.041\sigma^{-3}$ respectively, in very good agreement with 
the simulation values $\rho_{f}=0.945\sigma^{-3}$, 
$\rho_{m}=1.041\sigma^{-3}$. 
However, describing the hard-sphere fluid by the Percus-Yevick 
(PY) EOS~\cite{hansen} would fail to give a fluid-solid transition 
for this system, and consequently for the potential~(\ref{pot}) 
in the high-temperature limit. This consideration is relevant in the present
context, since our formulation of SCOZA represents the direct correlation
function $c(r)$ for $r>\sigma$ as the superposition of two Yukawa functions,
both of which are needed to represent the tail of the HCTY 
potential~(\ref{pot}). As a consequence, there should be no off-core 
contribution to $c(r)$ due to the purely hard-core part of $v(r)$, and this 
would indeed yield a PY description of the hard-sphere fluid. The conceptually 
most straightforward way to include the CS EOS into SCOZA is to resort 
to the Waisman parameterization~\cite{waisman} for the hard-sphere 
contribution to $c(r)$, so that the full $c(r)$ for $r>\sigma$ takes 
a three-Yukawa form. Such a treatment has been developed by other authors, 
and generalized to a linear combination of $n$ Yukawa tails~\cite{paschinger}
as well as to a linear combination of Yukawa and exponential 
tails~\cite{paschinger2}, albeit so far the applications do not include 
the mermaid potential considered here. For the sake of computational 
simplicity, we have instead resorted to an {\em ad hoc} modification 
of the two-Yukawa SCOZA, which consists in assuming that the hard-sphere 
contribution to $c(r)$ outside the repulsive core has a Yukawa form 
with the same 
inverse-length parameter $z_{1}$ as the shorter-ranged part of $v(r)$, and 
a density-dependent amplitude $H(\rho)$ which is set so as to give the CS EOS. 
As a consequence, in the full $c(r)$ the amplitude of the shorter-ranged 
Yukawa is given by $H(\rho)+K(\rho, \beta)$ rather than by 
$K(\rho, \beta)$ alone, where $H(\rho)$ is a known function, 
and $K(\rho,\beta)$ is determined by the thermodynamic consitency 
condition~(\ref{consist}) as specified above. 
This method has been 
illustrated in greater detail in reference~\cite{twoyuk}, where it was used 
to study a HCTY interaction with amplitudes and ranges fitted to 
a Lennard-Jones potential. The resulting liquid-vapour coexistence curve was 
found to be both accurate and very close to that subsequently 
obtained~\cite{paschinger3} for
the same interaction by the three-Yukawa formulation of SCOZA: the relative
difference in the critical points does not exceed $0.33\%$. Perhaps more 
surprisingly, also the radial distribution function appeared to be little 
affected by our {\em ad hoc} assumption on the hard-sphere contribution 
to $c(r)$~\cite{scozatwoyuk}.
To get an estimate of how much the location of the fluid-solid phase boundary
is affected by this prescription, we set $A=0$ in Eq.~(\ref{pot}), so 
as to get the same hard-core Yukawa (HCY) interaction that was studied 
in reference~\cite{foffi}. For this interaction, the Waisman description 
of the hard-sphere reference fluid is achieved within our two-Yukawa SCOZA, 
and in fact this is the approach that has been used in that work. We then 
employed the simplified treatment of the hard-sphere reference fluid adopted
here to determine the value of the inverse range $z_{1}$ at which 
the liquid-vapour transition becomes metastable with respect to the fluid-solid 
one, and compared it with that obtained in reference~\cite{foffi}. 
That calculation
gave $z_{1}=6.05$, in close agreement with the Gibbs ensemble Monte Carlo
(GEMC) result $z_{1}=6$ by Hagen and Frenkel~\cite{hagen}, while here we get 
$z_{1}=5.7$. We regard this $5\%$ discrepancy with respect to the more 
complete treatment as a tolerable one, especially in view of the fact that
the considerations developed in this work are mainly qualitative. 

\section{Phase diagram} 
\label{sec:freezing}
   
In the following we will adopt reduced quantities 
$T^{*}=k_{\rm B}T/\epsilon$, $\beta^{*}=\beta\epsilon$, 
$\rho^{*}=\rho\sigma^{3}$. The SCOZA PDE that we used for the fluid phase
has been solved numerically on a grid of $(\beta^{*}, \rho^{*})$ values. 
The reduced density interval extended from $\rho^{*}=0$ up to an high-density
boundary $\rho_{0}^{*}$. This was set at $\rho_{0}^{*}=1$ for potentials 
with relatively long-ranged attractive parts, e.g. for inverse-range parameter
$z_{1}=1.8$. For short-ranged interactions, the liquid-vapour coexistence
curve rapidly extends to high densities as the temperature is lowered,
and the high-density boundary was therefore moved to higher values, up to 
$\rho_{0}^{*}=1.4$ for the case $z_{1}=7$. 
The integration was stopped at a value 
of $\beta^{*}$ large enough to obtain a significant portion 
of the liquid-vapour coexistence curve. The pressure and the chemical potential
of the solid phase needed to determine the fluid-solid equilibrium lines were
obtained by numerical differentiation of the solid free energy given by
Eqs.~(\ref{solid}), (\ref{mc}). 

\begin{figure}
\centerline{\psfig{figure=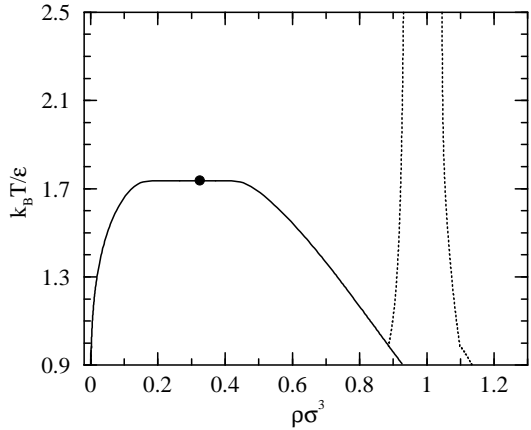,width=7cm}}
\caption{Phase diagram of the hard-core two-Yukawa fluid with competing 
interactions (see Eq.~(\protect\ref{pot})) in the density-temperature plane
according to the SCOZA supplemented by perturbation theory for the solid
phase (see text). 
The potential parameters are $z_{1}=1$, $z_{2}=0.5$, $A=0.0976$. 
Solid line: liquid-vapour coexistence. Dotted line: fluid-solid coexistence. 
Full circle: liquid-vapour critical point.}
\label{fig:z1}  
\end{figure}

In figure~\ref{fig:z1} we show the phase diagram corresponding to 
the inverse-range parameters $z_{1}=1$, $z_{2}=0.5$ that were already 
considered in reference~\cite{chemlet}. The amplitude of the repulsive tail
was set at the value $A=0.0976$, at the stability limit of bulk   
liquid-vapour transition. 
For slightly larger values of $A$, our numerical
algorithm fails to converge before the liquid-vapour critical point is reached,
and we take this as an indication that, as the temperature is lowered, 
the system would undergo a transition into a nonhomogeneous phase. This cannot
be described by a liquid-state theory such as the SCOZA, which assumes 
the fluid to be homogeneous. We observe that the boundary value of $A$ 
is little affected by the {\em ad hoc} procedure described 
in Sec.~\ref{sec:theory} 
to ensure that SCOZA recovers the CS hard-sphere EOS in the limit 
$\beta\rightarrow 0$. The relative difference with respect to the value 
obtained using PY hard-sphere thermodynamics 
as in reference~\cite{chemlet} does
not exceed $0.3\%$. For this choice of parameters, the system shows stable 
vapour, liquid, and solid phases with a triple point. This is in agreement 
with what we found in reference~\cite{chemlet} by means of the Hansen-Verlet 
criterion for freezing~\cite{verlet}. 
The phase diagram illustrated in figure~\ref{fig:z1} is representative 
of the situation that we found at small values of $z_{1}$. Specifically, 
we chose $z_{1}=1$ and $z_{1}=2$ and considered several values of $z_{2}$,
namely $z_{2}=0.25$, $0.5$, $0.75$, $0.9$ for $z_{1}=1$, and $z_{2}=1.5$,
$1.8$, $1.9$ for $z_{1}=2$. In all of these cases, we found that freezing 
never preempts the liquid-vapour transition in the whole range of repulsion 
strength $A$ at which the latter occurs. We recall that on increasing $A$ 
within this range, the critical temperature decreases. As we observed 
in reference~\cite{chemlet}, there is a trivial contribution to this effect 
which
is to be expected already at the mean-field level, just because the integrated
intensity of the (overall attractive) potential decreases in absolute value
as $A$ is increased. However, close to the boundary value of $A$, where 
the repulsion affects more deeply the behaviour of the fluid, the drop 
in the critical temperature becomes much sharper than that given by mean-field
theory. For a given value of $z_{1}$, the decrease in the critical temperature 
with respect to that of the purely attractive case gets larger as the inverse
range of the repulsion $z_{2}$ gets closer to $z_{1}$. Nevertheless, this does 
not imply that by taking the two ranges very similar, the liquid-vapour 
coexistence curve will be pushed below the freezing curve, thereby having
freezing preempt liquid-vapour phase separation. In fact, the limit 
$z_{1}=z_{2}$ amounts to a trivial rescaling of the interaction strength with 
respect to the purely attractive case $A=0$, which is not going to affect
the shape of the phase diagram, provided of course the sign of the potential
remains negative.  
  
\begin{figure}
\centerline{\psfig{figure=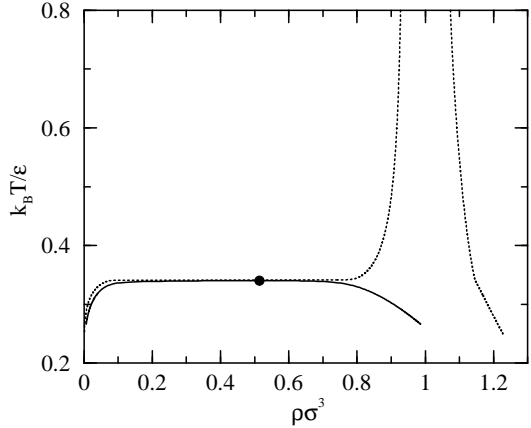,width=7cm}}
\caption{Same as figure~\protect\ref{fig:z1} with $z_{1}=3$, $z_{2}=1.8$, 
$A=0.29$.}
\label{fig:z3}
\end{figure}

We now consider larger values of $z_{1}$. Figure~\ref{fig:z3} shows the phase 
diagram for $z_{1}=3$, $z_{2}=1.8$, $A=0.29$. The situation has changed with 
respect to that illustrated in figure~\ref{fig:z1}, since here 
the liquid-vapour
coexistence curve is tangent to the freezing line, and disappears into 
the fluid-solid coexistence region for larger values of $A$, until the 
stability limit of bulk phase separation is reached. 
For the case shown in 
figure~\ref{fig:z3}, this happens for $A=0.295$, which gives a metastable 
liquid-vapour critical temperature differing from that of the flat part 
of the liquidus line by about $10\%$. A similar behaviour is found, 
for instance, for $z_{1}=4$, $z_{2}=1.8$, and $0.15\leq A\leq 0.17$, 
the largest relative difference in temperature between the metastable critical
point and the plateau of the liquidus line being now around $25\%$. We observe
that in this regime one finds a metastable liquid-vapour coexistence curve for
an inverse range $z_{1}$ of the attractive tail that is significantly smaller
than the threshold value for metastability with purely attractive 
interactions. The latter, as we said in Sec.~\ref{sec:theory}, amounts to 
$z_{1}=5.7$ in the approximation used here. It may then be worthwhile looking
more closely at the origin of the metastability in the present case. 
A possible explanation is that, as $A$ is increased, the repulsive part 
of the potential leads to a decrease in the effective range of the attractive
contribution, which in turn makes the liquid-vapour critical point become 
metastable, just as in fluids with short-ranged attractive interactions. 
However, this is not always the case: for instance, the profile
of the potential for the parameter choice $z_{1}=3$, $z_{2}=1.8$, $A=0.29$ 
to which figure~\ref{fig:z3} refers, is indeed narrower than a single 
attractive tail with $z_{1}=3$, but nevertheless it is quite wider than 
the tail corresponding to the threshold range $z_{1}=5.7$. 
A quantitative comparison can be made by means of the prescription
for determining the effective attraction range proposed by Noro 
and Frenkel~\cite{noro} in order to give a unified description of the phase
behaviour of fluids with different attractive potentials. For a given 
interaction, this temperature-dependent range is defined as the width of 
the square-well potential which yields the same reduced second virial 
coefficient $B_{2}(T)/B_{2}^{\rm HS}$ at the same reduced temperature 
$k_{\rm B}T/v(\sigma)$, where the reduced quantities are obtained by rescaling
the temperature $T$ and the second virial coefficient $B_{2}(T)$ by the depth
of the attractive well $v(\sigma)$ and by the hard-sphere second virial 
coefficent $B_{2}^{\rm HS}=2\pi\sigma^{3}/3$ respectively. If we adopt this
mapping for the case just mentioned, we find that at the critical temperature
the effective range of the interaction with $z_{1}=3$, $z_{2}=1.8$, $A=0.29$
is $40\%$ larger than that of the purely attractive interaction with 
$z_{1}=5.7$, $A=0$, despite the fact that both of them give a liquid-vapour
coexistence curve tangent to the freezing line. Therefore, trying to predict
the phase diagram of a fluid with competing interactions on the basis of 
the effective range alone may turn out to be tricky, even before bulk 
phase separation disappears. In fact, Noro and Frenkel~\cite{noro} observed 
that, unlike interactions whose off-core part is purely attractive, potentials
that include repulsive barriers do not lend themselves to a universal 
description in terms of their effective range. However, we note that, 
in the region where the liquidus line broadens, the rescaled temperature 
$k_{\rm B}T/v(\sigma)$ is little affected by the repulsion. This agrees with
what found by Louis~\cite{louis1,louis2} for fluids with deep attractive 
wells. For these ``energetic fluids'', he determined an approximate 
crystallization criterion~\cite{louis2}, according to which the freezing 
density becomes about $50\%$ of that of the hard-sphere fluid when one has 
$\beta v(\sigma)\simeq 2.4 \pm 0.3$, where $\beta v(\sigma)$ is the reciprocal 
of the rescaled temperature.
Specifically, for the above-mentioned cases $z_{1}=5.7$, $A=0$; $z_{1}=4$, 
$z_{2}=1.8$, $A=0.15$; $z_{1}=3$, $z_{2}=1.8$, $A=0.29$; we find that 
the liquidus line broadens to half the hard-sphere freezing density 
for $\beta v(\sigma)=2.15$, $\beta v(\sigma)=2.17$, and $\beta v(\sigma)=2.08$
respectively, in fair agreement with Louis' criterion. Such a result strongly
suggests that even in the regime to which figure~\ref{fig:z3} refers, where 
the liquid-vapour transition would be stable in the absence of repulsion, it is
still the effective strength $\beta v(\sigma)$ of the attraction at contact 
that causes the metastability of liquid-vapour phase separation. This occurs
when the inverse critical temperature $\beta_{c}$ is large enough for 
$\beta_{c} v(\sigma)$ to exceed the threshold value at which the liquidus line
broadens. Therefore, the change in the phase diagram from stable to metastable
liquid-vapour transition is triggered essentially by the same mechanism 
as in the case of narrow attractive potentials. The only difference is that 
here we have a sort of additional degree of freedom with respect to 
a purely attractive interaction, insomuch as the increase of the rescaled 
inverse critical temperature $\beta_{c} v(\sigma)$ beyond the threshold for
metastability can be induced not only by decreasing the attraction range, 
but also by increasing the strength of the repulsion for suitable choices 
of the inverse-range parameters $z_{1}$, $z_{2}$ so as to approach 
the boundary of bulk phase separation, where the critical temperature drops 
quickly. In the latter case, the liquid-vapour transition can become metastable
at potential ranges that are significantly larger than those required for 
purely attractive interactions. 

\begin{figure} 
\centerline{\psfig{figure=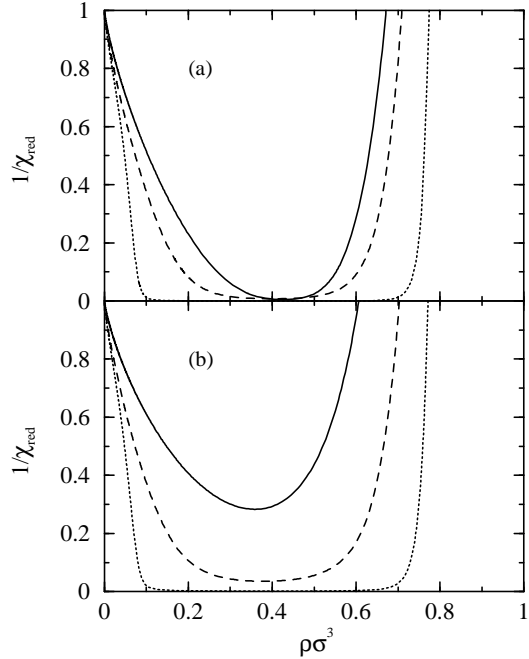,width=7cm}}
\caption{Inverse reduced compressibility $1/\chi_{\rm red}$ along the liquidus 
line for different interaction profiles. The interaction 
parameters have been chosen so that in panel~(a) the liquid-vapour coexistence 
curve is tangent to the liquidus line, while in panel~(b) it is metastable,
the flat part of the liquidus line corresponding to a temperature about $8\%$
above that of the liquid-vapour critical point.     
Panel~(a): solid line, $z_{1}=5.7$, $A=0$; dashed line, $z_{1}=4$, 
$z_{2}=1.8$, $A=0.15$; dotted line, $z{1}=3$, $z_{2}=1.8$, $A=0.29$.
Panel~(b): solid line, $z_{1}=7$, $A=0$; dashed line, $z_{1}=4$, $z_{2}=1.8$,
$A=0.16$; dotted line, $z_{1}=3$, $z_{2}=1.8$, $A=0.294$.}
\label{fig:comp}
\end{figure}

Moreover, the proximity to the stability limit
for the existence of the homogeneous fluid phases implies that the effects
on the thermodynamics which are characteristic 
of that regime, summarized in Sec.~\ref{sec:introduction}, show up along 
and in the neighborhood of the liquidus line. For instance, 
in figure~\ref{fig:z3} this line shares with the liquid-vapour coexistence curve
the very flat portion that develops in the latter shortly before the repulsion
amplitude $A$ reaches its boundary value. As a consequence, on lowering 
the temperature the liquidus line broadens even more abruptly than in fluids 
with narrow attractive interactions. The effect on the density fluctuations 
is shown in figure~\ref{fig:comp}, that compares the reciprocal of the reduced
compressibility $1/\chi_{\rm red}$ along the liquidus line for a purely 
attractive tail interaction and for the cases $z_{1}=3$, $z_{2}=1.8$ and 
$z_{1}=4$, $z_{2}=1.8$. 
In panel (a), the inverse range of the attractive 
interaction and the amplitudes of the repulsive tail have been chosen so that
the liquidus line is tangent to the liquid-vapour coexistence curve. 
While the compressibility obviously always increases as the critical point 
is approached, the size of the region where it remains large is greatly 
enhanced by the competition between attraction and repulsion. This region 
grows up not only in density but also in temperature, as shown in panel (b), 
where the interaction parameters have been adjusted so that the flat part 
of the liquidus line lies above the metastable liquid-vapour critical 
temperature by about $8\%$. As the liquid-vapour coexistence curve sinks below
the liquidus line, the $1/\chi_{\rm red}$ vs. $\rho$ curve corresponding to 
the purely attractive potential is pulled upward much more quickly than those
of the fluids with competing interactions. 
In summary, 
in the intermediate interval of $z_{1}$ just considered, the strong density 
fluctuations induced by the competing interactions markedly affect 
the behaviour of the fluid in the region of the liquidus line, even though
the metastability of the liquid-vapour transition is always driven 
by the strength of the attraction near contact.  
 
\begin{figure}
\centerline{\psfig{figure=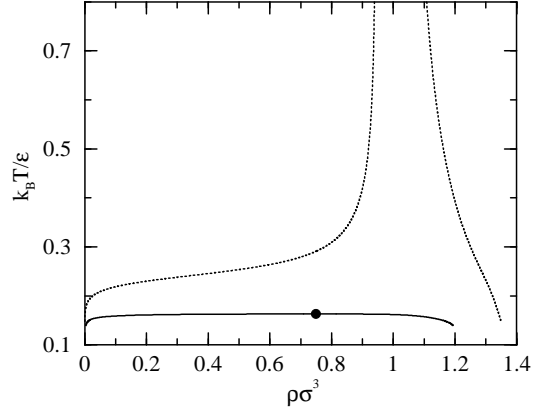,width=7cm}}
\caption{Same as figure~\protect\ref{fig:z1} with $z_{1}=7$, $z_{2}=4$, 
$A=0.4$.}
\label{fig:z7}
\end{figure}

Finally, we turn to the case where the inverse range $z_{1}$ of the attractive 
part of the interaction is large enough to give a metastable liquid-vapour 
transition even in the absence of repulsion. In this regime, the liquid-vapour
transition remains {\em a fortiori} metastable when the repulsion is 
turned on. As $A$ is increased, the temperature of the metastable critical
temperature decreases more quickly than that of the plateau of the liquidus
line, so that the liquid-vapour coexistence curve never gets out 
of the fluid-solid coexistence region. Therefore, the effects 
on the thermodynamics and correlations that come along because of 
the competing interactions always involve the domain of the metastable fluid.
An example of the phase diagram corresponding to this situation is shown 
in figure~\ref{fig:z7} for $z_{1}=7$, $z_{2}=4$, $A=0.4$. The metastable 
liquid-vapour coexistence curve exhibits the very flat shape that one finds 
close to the boundary value of $A$ ($A=0.42$ in the present case), 
beyond which no bulk fluid-fluid phase separation is predicted by the theory. 

A quite relevant issue 
is the role played by the freezing transition when the strength 
of the repulsion becomes large enough to give rise to the formation 
of inhomogeneous phases. Specifically, one might ask whether in this regime
of short-ranged attraction the freezing transition will preempt the occurrence
of microphases just as it preempts the liquid-vapour transition at lower values
of $A$. A fully satisfactory answer cannot be given within the present
investigation, since, as we noted above, our theoretical approach cannot deal
with nonuniform phases. However, we can try and increase $A$ enough to get 
into the region of interaction parameters where the fluid is expected 
to become inhomogeneous at low enough temperature. We then find that 
the temperature at which our algorithm does not yield solutions rapidly 
increases with $A$. For the inverse-range choice $z_{1}=7$, $z_{2}=4$ to which
figure~\ref{fig:z7} refers, it is sufficent to raise the repulsion strength 
slightly above $A=0.5$ to find that at the lowest isotherm that we can attain,
the freezing line has not developed a plateau yet.      
This suggests that for high enough $A$, the formation of microphases will not
be preempted by freezing, even when the attractive part of the interaction 
is short-ranged. The fact that the temperature of the transition from 
the homogeneous to the density-modulated phase should increase as 
the repulsion strength is increased is not surprising, and is
in agreement with the results obtained in lattice systems with competing 
interactions~\cite{lattice}. 

\section{Correlations}
\label{sec:correlations}

As already observed in reference~\cite{chemlet}, competing interactions deeply 
affect the correlations, even in the region where the system is still 
in the fluid phase. In the OZ picture, the behaviour of the correlations 
in the region where the compressibility is large is obtained by assuming
a parabolic shape for the reciprocal of the structure factor $S(q)$ at small
wavevector $q$, $1/S(q)\sim a+b q^{2}$ with $a=1/\chi_{\rm red}$. 
This gives
\begin{equation}
h(r)\sim \frac{1}{4\pi \rho \, b}\left(1-\frac{1}{\, \chi_{\rm red}}\right)
\frac{1}{r} \, \exp(-r/\xi)
\label{oz}
\end{equation}
for the asymptotic decay of the two-body correlation function $h(r)$, where
$\xi=\sqrt{b/a}$ is the usual correlation length. 
As is well known, the assumption that $1/S(q)$ is analytic in $q$ breaks down
at the critical point, but even there it is still a reasonable approximation,
because in three dimensional fluids the deviations from analyticity are small. 
However, when the strength $A$ of the repulsive
part of the potential is large enough for the system to be close to 
the stability limit of bulk liquid-vapour phase separation, the coefficient $b$
can be very small. The link between $b$ 
and the interaction is particularly straightforward within the random-phase
approximation (RPA),  
where this coefficient is proportional to the second moment
of the tail potential. In the RPA scenario, the limit of bulk phase
separation is indeed reached at that value of $A$ for which $b$ vanishes. 
For larger values of $A$, $b$ becomes negative, so that $S(q)$ changes
its convexity at $q=0$. The peak of $S(q)$ then moves away from the origin,
and eventually develops into a singularity at nonzero $q$ at low temperature.
In more sophisticated approaches, such as the SCOZA used here, the explicit
form of $b$ cannot be worked out so easily, since this involves 
the state-dependent amplitude $K(\rho,\beta)$, which is {\em a priori} 
unknown, and is also expected to be a functional of the interaction itself. 
Nevertheless, it is still true that increasing the repulsion strength $A$ 
leads in general to a decrease and eventually to a negative value of $b$. 
For large enough $A$ the SCOZA structure factor displays,
in addition to the usual particle-particle peak at $q\simeq 2\pi/\sigma$,
the characteristic peak at smaller $q$ related to cluster-cluster 
correlations~\cite{stradner,stiakakis,sciortino}. However, as the temperature 
is lowered, the theory eventually fails to converge since, as observed 
in section~\ref{sec:freezing}, the SCOZA cannot describe 
the formation of inhomogeneous microphases. 

\begin{figure}
\centerline{\psfig{figure=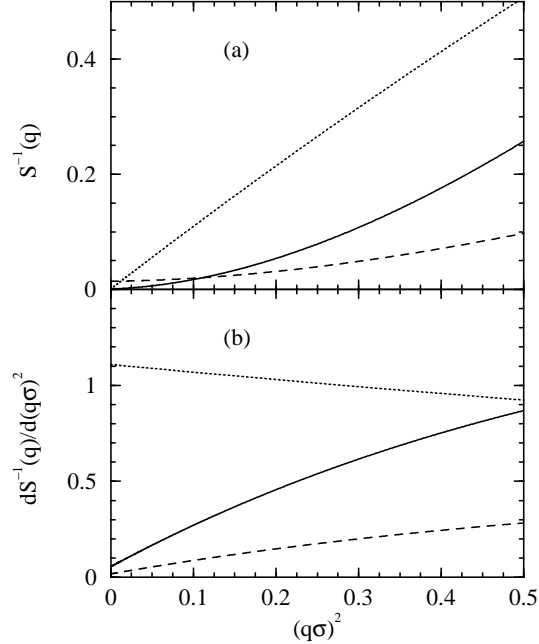,width=7cm}}
\caption{Panel (a): reciprocal of the structure factor $1/S(q)$ as a function 
of $q^{2}$ with and without competing interactions for the potential 
parameters and the thermodynamic states described in the text. 
Dotted line: $A=0$, $z_{1}=3$, $\rho^{*}\!=\!\rho^{*}_{c}=0.359$, 
$t=(T-T_{c})/T_{c}=6.45\times 10^{-4}$. Solid line: $A=0.29$, $z_{1}=3$, 
$z_{2}=1.8$, $\rho^{*}\!=\!\rho^{*}_{c}=0.515$, $t=2.81\times 10^{-2}$. 
Dashed line:
$A=0.29$, $z_{1}=3$, $z_{2}=1.8$, $\rho^{*}=0.1$, $t=3.5\times 10^{-3}$.
Note the deviations from linearity when competing interactions 
are present. Panel (b): derivative of $1/S(q)$ with respect to $q^{2}$ 
as a function of $q^{2}$. Eq.~(\protect\ref{quartic}) gives a straight line 
on this plot.}
\label{fig:oz}
\end{figure}

Even in the regime where the liquid-vapour transition still exists but 
the competition between attraction and repulsion is strong, 
the quadratic approximation for $1/S(q)$ at small $q$ is expected
to become a poor one, since higher-order terms become important. 
This is shown in figure~\ref{fig:oz}. 
In panel (a) we have plotted the SCOZA results for $1/S(q)$ as 
a function of $q^{2}$ for two fluids with and without competing interactions,
namely for $A=0.29$, $z_{1}=3$, $z_{2}=1.8$ and for $A=0$, $z_{1}=3$. 
In the latter case, we have $\rho^{*}\!=\!\rho^{*}_{c}=0.359$, 
$t=(T-T_{c})/T_{c}=6.45\times 10^{-4}$. 
For the competing interaction case, the density has again been set to 
the critical value $\rho^{*}_{c}=0.515$ and the reduced temperature is 
$t=2.81\times 10^{-2}$. The reduced compressibilities are similar for the
two states, $\chi_{\rm red}=1.61\times 10^{3}$ and 
$\chi_{\rm red}=1.67\times 10^{3}$ respectively. While for purely attractive
interactions $1/S(q)$ is a nearly linear function of $q^{2}$ in the interval
of $q$ reported in the figure, there are marked deviations from linearity 
when competing interactions are present. These deviations persist even 
at densities away from the critical one, as shown by the result for 
$\rho^{*}=0.1$, $t=3.50\times 10^{-3}$, $\chi_{\rm red}=71.3$, also plotted 
in the figure, which lies very close to the liquidus line.
One is then lead to take into account also   
the $q^{4}$-term in the expansion of $1/S(q)$ by setting 
\begin{equation}
1/S(q)\sim a+b q^{2}+c q^{4} \, .
\label{quartic}
\end{equation}
To which extent this can be considered
a good approximation is shown in panel (b) of figure~\ref{fig:oz},
which shows the derivative of $1/S(q)$ with respect to $q^{2}$ as a function 
of $q^{2}$. In this representation, Eq.~(\ref{quartic}) gives 
a straight line with an intercept $b$ and a slope $2c$, the standard OZ form
corresponding to a horizontal line. It appears that in the fluid with  
competing interactions the $q^{4}$-term plays a much more important role than 
in the purely attractive case. Even for the low-density state   
$\rho^{*}=0.1$, $t=3.50\times 10^{-3}$ for which the absolute value of the 
deviation from the OZ form is similar to that of the attractive case, 
the relative weight of the quartic term is nevertheless much bigger because
of the much smaller value of $b$.   

\begin{figure}
\centerline{\psfig{figure=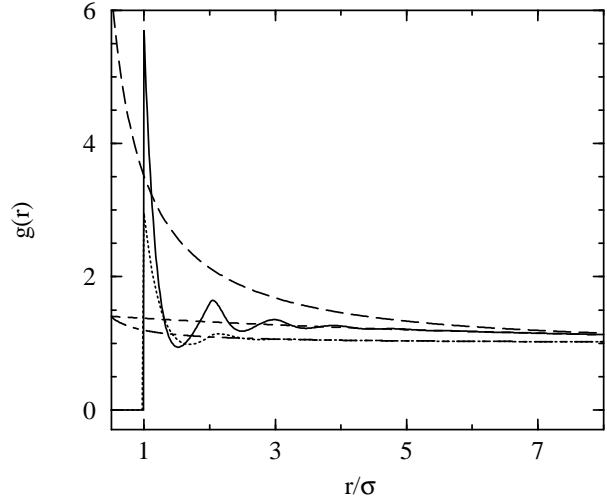,width=8cm}}
\caption{Radial distribution function $g(r)$ at the critical density 
for the potential parameters
and thermodynamic states considered in figure~\protect\ref{fig:oz}. 
Solid line, $A=0.29$, $z_{1}=3$, $z_{2}=1.8$, $\rho^{*}=0.515$, 
$t=2.81\times 10^{-2}$, $\chi_{\rm red}=1.67\times 10^{3}$. 
Long-dashed line: long-range tail of $g(r)$ as predicted by the
OZ Eq.~(\protect\ref{oz}). Dashed line: long-range tail of $g(r)$ as predicted
by the double-exponential Eq.~(\protect\ref{h1}). Dotted line: $A=0$, 
$z_{1}=3$, $\rho^{*}=0.359$, $t=6.45\times 10^{-4}$, 
$\chi_{\rm red}=1.61\times 10^{3}$. Dot-dashed line: long-range 
tail of $g(r)$ as predicted by the OZ Eq.~(\protect\ref{oz}).}
\label{fig:corrlong}
\end{figure}

The long-range behaviour of the correlation function corresponding 
to Eq.~(\ref{quartic}) is obtained by 
a straightforward contour integration, and has the form~\cite{chemlet}
\begin{equation}
h(r)=\frac{1}{4\pi \rho \sqrt{\Delta}}
\left(1-\frac{1}{\, \chi_{\rm red}}\right)\frac{1}{r} 
\left[\, \exp(-r/\lambda)-\exp(-r/\mu)\, \right] \, ,
\label{h1}
\end{equation}
where $\Delta=b^{2}-4ac$, and the lengths $\lambda$, $\mu$ are given by 
\begin{eqnarray}
\lambda & = & \left(\frac{2c}{b-\sqrt{\Delta}}\right)^{\frac{1}{2}} \, ,
\label{lambda} \\
\mu & = & \left(\frac{2c}{b+\sqrt{\Delta}}\right)^{\frac{1}{2}} \, .
\label{mu}
\end{eqnarray}
In figure~\ref{fig:corrlong} we have plotted the radial distribution function
$g(r)=1+h(r)$ for the same interactions and thermodynamic states considered
in figure~\ref{fig:oz}. 
Unlike in the purely attractive case, the 
two-exponential form is indeed necessary to describe the long-range 
decay of the correlations. The usual OZ form~(\ref{oz}) is recovered only
asymptotically close to the critical point, where $\lambda$ diverges and 
reduces to the usual correlation length $\xi$, while $\mu$ saturates 
at a finite value. This does not apply to the situation shown in the figure,
for which we find $\lambda=7.80$, $\mu=5.65$ in units of the hard-sphere 
diameter $\sigma$. Besides being comparable, these lengths are much smaller 
than the correlation length $\xi=42.3\, \sigma$ of the purely attractive case. 
Since the compressibility is similar for the two states considered here, 
the slower decay of $h(r)$ in the attractive fluid must be compensated
by larger values of $h(r)$ at small and intermediate $r$ in the fluid 
with competing interacions. This is the reason for the enhancement 
of the correlations induced by the competition in the range of $r$ shown
in the figure. 

Eqs.~(\ref{h1})--(\ref{mu}) are formally valid also for 
$b<2\sqrt{ac}$, but they do not appear physically transparent in this case,
since $\lambda$ and $\mu$ become complex conjugate quantities, and $h(r)$
presents damped oscillations at large $r$. In this regime, 
Eqs.~(\ref{h1})--(\ref{mu}) are more conveniently rewritten as follows:    
\begin{equation}
h(r)=\frac{1}{2\pi \rho \sqrt{|\Delta|}}
\left(1-\frac{1}{\, \chi_{\rm red}}\right)\frac{1}{r} \, 
\sin(r/\delta) \exp(-r/\zeta) \, ,
\label{h2}
\end{equation}
\begin{eqnarray}
\delta & = & 2\left(\frac{c}{2\sqrt{ac}-b}\right)^{\frac{1}{2}}
\, ,
\label{delta}  \\
\zeta & = & 2\left(\frac{c}{2\sqrt{ac}+b}\right)^{\frac{1}{2}}
\, .
\label{zeta} 
\end{eqnarray}
It is interesting to observe that 
the asymptotic decay of $g(r)$ can be oscillatory even for states where 
the compressibility is large.   
The occurrence of monotonic and oscillatory regimes for the decay 
of the correlations, governed respectively by the characteristic lengths 
(\ref{lambda}), (\ref{mu}) and (\ref{delta}), (\ref{zeta}), was obtained 
before by other authors~\cite{nussinov} in a general study of $O(n)$ 
spin models with competing ferro- and antiferromagnetic interactions.
In fact, the effects discussed above stem from rather general features 
of the interaction and appear close to the limit of bulk 
fluid-fluid phase separation irrespective of the specific choice 
of the competing-interaction potential, provided of course the 
amplitude $A$ of the repulsion is tuned so that one is indeed close 
to that limit. 

\begin{figure}
\centerline{\psfig{figure=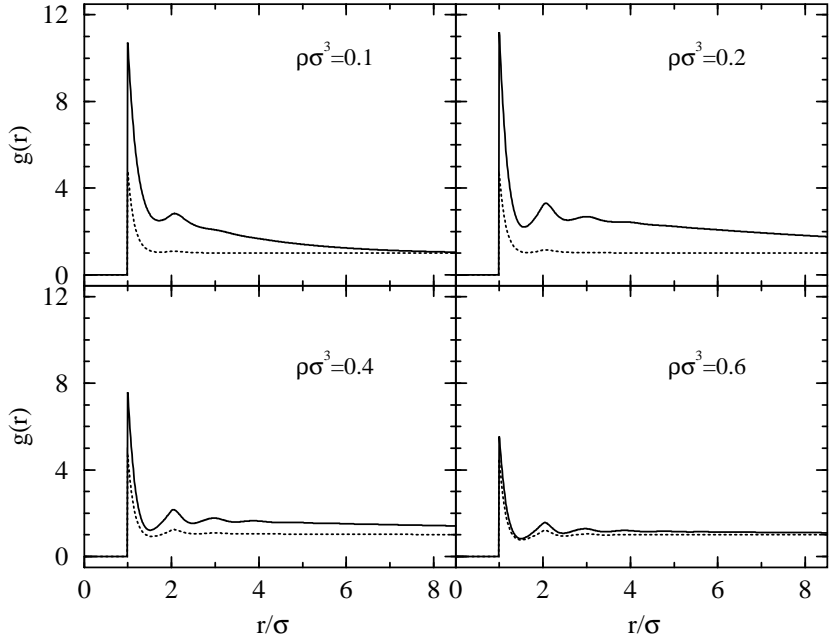,width=11cm}}
\caption{Radial distribution function $g(r)$ along the liquidus line for 
$z_{1}=5.7$, $A=0$ (dotted line) and $z_{1}=3$, $z_{2}=1.8$, $A=0.29$ 
(solid line). Each panel corresponds to a different density.} 
\label{fig:corr}
\end{figure}

They can be relevant also for the correlations  
along the liquidus line, if the interaction parameters are such that
this line enters the large-fluctuation region as it broadens 
to low density. One then expects the behaviour of the correlations
to be different from that found along the liquidus line of purely
attractive fluids.  
The two situations are compared 
in figure~\ref{fig:corr}, where $g(r)$ is shown at several densities along 
the liquidus line for $z_{1}=5.7$, $A=0$ and $z_{1}=3$, $z_{2}=1.8$, $A=0.29$. 
For both of these sets of parameters, as observed above, the liquidus line 
is tangent to the liquid-vapour coexistence curve. 
For systems with narrow 
attractions, it has been observed by Louis~\cite{louis2} that, in the region
where the liquidus line broadens, the two-body radial 
distribution function $g(r)$ is well approximated by the zeroth-order term 
of its expansion in powers of the density, $g(r)\simeq \exp[-\beta v(r)]$, 
in a remarkably wide density interval. As a consequence, varying the density
within this interval has little effect on $g(r)$. 
While the $g(r)$ 
corresponding to the purely attractive tail is indeed relatively unaffected 
by density, that of the fluid with competing interactions deviates almost 
immediately from its low-density limit, and is strongly enhanced both near
contact and at larger values of $r$ as soon as large density fluctuations 
develop. We recall that in SCOZA the direct
correlation function is linear in the off-core part of the interaction, as in
the MSA. As a consequence, the $g(r)$ obtained by the theory does not 
reproduce the exact zeroth-order term of the density expansion, and shows 
some quantitative inaccuracies, especially near contact, where it 
underestimates the correct values. Nevertheless, we expect the qualitative
behaviour shown in figure~\ref{fig:corr} to be genuine.

\section{Conclusions}
\label{sec:conclusions}

We have used SCOZA to study the correlations in fluids with competing
interactions consisting of the sum of an attractive Yukawa tail and
a longer-ranged, repulsive one. Our results show that 
the competition between attraction and repulsion causes the appearance of
strong density fluctuations in a relatively large region 
of the temperature-density plane. In this regime, the usual OZ picture, 
according to which the correlations in the critical region are described
in terms of a single correlation length, is replaced by a scenario where 
two characteristic lengths are necessary to describe the long-range
behaviour of the correlations. 
We have also considered the freezing transition and the stability 
of liquid-vapour phase separation with respect to freezing by supplementing  
SCOZA with thermodynamic perturbation theory for the solid. 
We find that in most cases the character of the liquid-vapour transition 
is determined by the range of the attractive contribution: potentials
with long- or short-range attractive parts give a liquid-vapour critical point
which lies respectively outside and inside the fluid-solid coexistence region.
However, at intermediate attraction ranges one can have freezing preemept 
the liquid-vapour transition by suitably tuning the repulsive contribution,
and the region of large fluctuations encompasses the liquidus line.  
It would be interesting to find out if this effect can be relevant 
for nucleation, especially of the solid phase in the fluid~\cite{frenkel}.  
We finally remark that 
we focused on a two-Yukawa potential because it leads to a semi-analytic 
formulation of SCOZA. However, the effects discussed here appear 
to hinge on quite general features of the competition between attraction 
and repulsion as the stability limit of bulk fluid-fluid
phase separation is approached. 
Therefore, they are expected to occur also in different systems with competing
interactions~\cite{nussinov}.

\ack
We acknowledge support from the Marie Curie program of the
European Union, contract number MRTN-CT2003-504712.


\section*{References}

\begin{thebibliography}{99}
%
\bibitem{brazovskii} S~A~Brazovski\v{\i}, {\it Zh. Eksp. Teor. Fiz.}
{\bf 68}, 175 (1975) [{\it Sov. Phys. JETP} {\bf 41}, 85 (1975)].
%
\bibitem{andelman} D~Andelman, F~Bro\c{c}hard, and J~F~Joanny,
{\it J. Chem. Phys} {\bf 86}, 3673 (1987). 
%
\bibitem{seul} M~Seul and D~Andelman, {\it Science} {\bf 267}, 476 (1995).
%                                                                     
\bibitem{gelbart1} W~M~Gelbart, R~P~Sear, J~R~Heath, and
S~Chaney, {\it Faraday Discuss.} {\bf 112}, 299 (1999).   
%
\bibitem{gelbart2} R~P~Sear, S~W~Chung, G~Markovich,
W~M~Gelbart, and J~R~Heath, {\it Phys. Rev. E} {\bf 59}, R6255 (1999).   
%
\bibitem{imperio} A~Imperio and L~Reatto, {\it J. Phys.: Condens. Matter} 
{\bf 16}, S3769 (2004).
%
\bibitem{bartlett} A~I~Campbell, V~J~Anderson, J~S~van~Duijneveldt, 
and P~Bartlett, {\it Phys. Rev. Lett.} {\bf 94}, 208301 (2005).
%
\bibitem{stradner} A~Stradner, H~Sedgwick, F~Cardinaux, W~C~K~Poon,
S~U~Egelhaaf, and P~Schurtenberger, {\it Nature} {\bf 432}, 492 (2004).
%
\bibitem{stiakakis} E~Stiakakis, G~Petekidis, D~Vlassopoulos, C~N~Likos, 
H~Iatrou, N~Hadjichristidis, and J~Roovers, {\it Europhys. Lett.} {\bf 72}, 
664 (2005).
%
\bibitem{sciortino} F~Sciortino, S~Mossa, E~Zaccarelli, and P~Tartaglia,
{\it Phys. Rev. Lett.} {\bf 93}, 055701 (2004).
%
\bibitem{chemlet} D~Pini, G~Jialin, A~Parola, and L~Reatto, {\it Chem. 
Phys. Lett.} {\bf 327}, 209 (2000). 
%
\bibitem{verlet} J~P~Hansen and L~Verlet, {\it Phys. Rev.} {\bf 184}, 151
(1969).
%
\bibitem{giaquinta} P~V~Giaquinta and G~Giunta, {\it Physica A} {\bf 187},
145 (1992).
%    
\bibitem{scozatwoyuk} D~Pini, G~Stell, and N~B~Wilding, 
{\it J. Chem. Phys} {\bf 115}, 2702 (2001).   
%
\bibitem{hansen} See, for instance, J~P~Hansen and I~R~McDonald, 
{\it Theory of Simple Liquids} (Academic Press, London, 1986).  
%
\bibitem{twoyuk} J~S~H\o ye, G~Stell, and E~Waisman, {\it Mol. Phys.}
{\bf 32}, 209 (1976);
J~ S~H\o ye and G~Stell, {\it ibid.} {\bf 52}, 1057 (1984); 
{\bf 52}, 1071 (1984).  
%
\bibitem{barker} J~A~Barker and D~Henderson, {\it J. Chem. Phys.} {\bf 47},
2856 (1967).
%
\bibitem{weis} J~J~Weis, {\it Mol. Phys.} {\bf 28}, 187 (1974);
C~E~Hecht and J~Lind, {\it J. Chem. Phys.} {\bf 64}, 641 (1976);
J~M~Kincaid, G~Stell, and E~Goldmark, {\it J. Chem. Phys.} {\bf 65}, 
2172 (1976). 
%
\bibitem{gast} A~P~Gast, W~B~Russell, and C~K~Hall, {\it J. Colloid
Interface Sci.} {\bf 96}, 1977 (1983); {\bf 109}, 161 (1986). 
%
\bibitem{hagen} M~H~J~Hagen and D~Frenkel, {\it J. Chem. Phys.} {\bf 101}, 
4093 (1994). 
%
\bibitem{dijkstra} M~Dijkstra, J~M~Brader, and R~Evans, {\it J. Phys.: 
Condens. Matter} {\bf 11}, 10079 (1999). 
%
\bibitem{foffi} G~Foffi, G~D~McCullagh, A~Lawlor, E~Zaccarelli, 
K~A~Dawson, F~Sciortino, P~Tartaglia, D~Pini, and G~Stell,
{\it Phys. Rev. E} {\bf 65}, 031407 (2002).  
%
\bibitem{hall} K~R~Hall, {\it J. Chem. Phys.} {\bf 57}, 2252 (1972). 
%
\bibitem{kincaid} J~M~Kincaid and J~J~Weis, {\it Mol. Phys.} {\bf 34},
931 (1977).
%
\bibitem{hirschfelder} J~O~Hirschfelder, C~F~Curtis, and R~B~Bird,
{\it The Molecular Theory of Gases and Liquids} (John Wiley \& Sons, New
York, 1954). 
%
\bibitem{waisman} E~Waisman, {\it Mol. Phys.} {\bf 25}, 45 (1973).
%
\bibitem{paschinger} G~Kahl, E~Sch\"oll-Paschinger, and G~Stell,
{\it J. Phys.: Condens. Matter} {\bf 14}, 9153 (2002); E~Sch\"oll-Paschinger,
G~Kahl, {\it Europhys. Lett.} {\bf 63}, 538 (2003); 
D~Costa, G~Pellicane, C~Caccamo, E~Sch\"oll-Paschinger, and G~Kahl,
{\it Phys. Rev. E} {\bf 68}, 21104 (2003).
%
\bibitem{paschinger2} E~Sch\"oll-Paschinger, {\it J. Chem. Phys.} {\bf 120},
11698 (2004).
%
\bibitem{paschinger3} E~Sch\"oll-Paschinger, Doctoral Thesis, University of
Wien, 2002. 
%
\bibitem{noro} M~G~Noro and D~Frenkel, {\it J. Chem. Phys} {\bf 113}, 
2941 (2000). 
%
\bibitem{louis1} A~A~Louis, R~Finken, and J~P~Hansen, {\it Phys. Rev. E}
{\bf 61}, R1028 (2000). 
%
\bibitem{louis2} A~A~Louis, {\it Phil. Trans. R. Soc. Lond. A} {\bf 359},
939 (2001). 
%
\bibitem{lattice} W~Selke, in: C~Domb and J~L~Lebowitz (Eds.),
{\it Phase Transition and Critical Phenomena}, Vol.~15, Academic Press,
New York, 1992.     
%
\bibitem{nussinov} Z~Nussinov, J~Rudnick, S~A~Kivelson and L~Chayes  
{\it Phys. Rev. Lett.} {\bf 83}, 472 (1999).
%
\bibitem{frenkel} P~R~ten~Wolde and D~Frenkel, {\it Science}
{\bf 277}, 1975 (1997).   

\end{thebibliography}
\end{document}